\documentclass{aastex}
\usepackage{emulateapj5} 

\received{August 10 2002}
\accepted{October 11 2002}

\begin{document}

\title{Can radiative cooling and 
nongravitational heating explain simultaneously 
the global X-ray properties of clusters and 
the unresolved cosmic X-ray background ?}

\author{Yan-Jie Xue and Xiang-Ping Wu}

\affil{National Astronomical Observatories, Chinese Academy
                 of Sciences, Beijing 100012, China}

\begin{abstract}
Using a simple analytic approach we address the question of whether
radiative cooling, nongravitational heating and cooling plus heating 
models can simultaneously explain the observed global X-ray
properties (entropy and X-ray luminosity distributions) of 
groups and clusters and the residual soft X-ray background (XRB)
after discrete sources are removed. 
Within the framework of typical cold dark matter structure
formation characterized by an amplitude of matter power
spectrum $\sigma_8=0.9$,   
it is argued that while radiative cooling alone is able to 
marginally reproduce the entropy floor detected 
in the central regions of groups and clusters, it is insufficient 
to account for the steepening of the X-ray luminosity - temperature 
relation for groups and the unresolved soft XRB. 
A phenomenological preheating model, in which either 
an extra specific energy budget or an entropy floor is added 
to the hot gas in groups and clusters, fails in the recovery 
of at least one of the X-ray observed features.  
Finally, the soft XRB predicted by our combined model of cooling 
plus heating exceeds the observational upper limits  
by a factor of $\sim2$, if the model is required 
to reproduce the observed entropy and X-ray 
luminosity - temperature relationships of groups and clusters.
Inclusion of the cosmic variation of metallicity and the self-absorption of 
the cooled gas as a result of radiative cooling in groups and clusters, 
or exclusion of the contribution of nearby, massive clusters to
the XRB does not significantly alter the situation.
If the discrepancy is not a result of the oversimplification of 
our analytic models, this implies that either our current understanding 
of the physical processes of the hot gas is still incomplete, or
the normalization of the present power spectrum 
has been systematically overestimated. For the latter, 
both the X-ray properties of groups and clusters and the XRB 
predicted by preheating model and cooling plus heating model 
can be reconciled with the X-ray observations if a lower value of 
the normalization parameter $\sigma_8\approx0.7$ is assumed.
\end{abstract}


\section{Introduction}

A substantial fraction of the baryons in the local universe exists 
in the form of diffuse warm-hot intergalactic medium (IGM) 
with temperatures of $T\sim10^5$--$10^7$ K as a result of gravitationally
driven shocks and adiabatic compression as they fall onto large-scale
structures and collapsed dark halos (Cen \& Ostriker 1999). 
In the former case, the warm IGM may escape direct detection, but
measurements of the strength and power spectra of the cosmic soft X-ray 
background (XRB) (e.g. Soltan, Freyberg \& Tr\"umper 2001) and 
the Sunyaev-Zel'dovich effect (e.g. Bond et al. 2002) 
can set stringent constraints on its amount and distribution.
For the latter, the hot IGM in virialized dark halos such as groups and
clusters manifests itself as strong X-ray emission sources through 
thermal bremsstrahlung, which provides a powerful tool for studying 
not only the physical processes of the hot IGM but also the formation 
and evolution of groups and clusters.

While gravity and thermal pressure play potentially important roles in the 
overall distribution and evolution of the hot IGM in groups and 
clusters, the IGM also suffers from the influence of nongravitational 
effects such as radiative cooling, heating by supernovae and/or AGNs, 
nonthermal pressure, etc. In low-mass groups and the central regions
of clusters, nongravitational effect may even become dominant for
the evolution of IGM. 
Indeed, there is growing observational evidence for the 
presence of nongravitational effects in groups and clusters.
Among many arguments, the most convincing observational facts are 
the significant departure of the observed X-ray 
luminosity ($L_{\rm X}$) - temperature ($T$) relation of groups and clusters 
($L_{\rm X}\propto T^{3-5}$) from the prediction ($L_{\rm X}\propto T^2$)
of self-similar model (e.g. Edge \& Stewart 1991; 
David et al. 1993; Wu, Xue \& Fang 1999; 
Helsdon \& Ponman 2000; Xue \& Wu 2000 and references therein)
and the entropy ($S$) excess in the central cores of 
groups and clusters (Ponman, Cannon \& Navarro 1999; Lloyd-Davies,
Ponman \& Cannon 2000).  Two prevailing scenarios suggested thus far,
preheating and radiative cooling of the hot IGM, both of which 
tend to suppress 
the X-ray emission of the hot IGM heated by purely gravitational shocks 
and compression, have been shown to be indistinguishable in the 
explanation of the observed X-ray properties of groups and clusters
(Voit \& Bryan 2001b; Voit et al. 2002; 
Borgani et al. 2002 and references therein), 
although each model still has its own problem. For instance, preheating 
model suffers from the so-called energy crisis that an unreasonably
high efficiency of energy injection into the IGM from supernovae
must be required in order to bring the IGM to the energy level seen
in the $L_{\rm X}$-$T$ and $S$-$T$ relations of groups and clusters 
(Wu, Fabian \& Nulsen 1998, 2000; Tozzi 2001; etc.),
although energy supply by AGNs may help to reduce the discrepancy 
(Valageas \& Silk 1999). Radiative cooling model suffers from the 
overcooling problem, i.e., the material that has cooled out of 
the hot IGM greatly exceeds the observational limits (Balogh et al. 2001).
Moreover, cooling efficiency is also a major concern in the
explanation of the X-ray observed properties of groups and clusters
(Bower et al. 2001). Nonetheless, both (pre)heating and cooling are
two natural processes during the formation of galaxies, which is
justified by the measurements of the IGM enrichment at high redshifts 
and the stellar mass fraction ($\sim10\%$) in the local universe. 
Indeed, it has been realized recently that only can a combination of 
preheating by supernovae and radiative cooling of the IGM  
reproduce the observed X-ray properties of groups and clusters 
(Voit et al. 2002; Borgani et al. 2002).

An independent and sensitive probe of the physical properties of 
the IGM is through study of the soft XRB. It has been shown that 
the XRB produced by the gravitationally heated and 
bound IGM in groups and clusters within the standard framework of
hierarchical formation of structures vastly exceeds the upper limits 
set by current X-ray observations, lending further support  to
nongravitational heating scenario (Pen 1999; Wu, Fabian \& Nulsen 2001;
Bryan \& Voit 2001). Yet, it is also noted that some hydrodynamical 
simulations have yielded a soft XRB compatible with
current limits without inclusion of nongravitational heating
(Croft et al. 2001; Dav\'e et al. 2001; Phillips, Ostriker \& Cen 2001).
This partially reflects the difficulty in handling the 
complex processes of IGM evolution in groups and clusters. 
Essentially, an excess energy of about 1 keV/particle should be 
injected into the IGM in order to reduce the predicted soft XRB to 
a reasonable level. On the other hand, it is unlikely that 
radiative cooling alone is sufficient to eliminate the
discrepancy (Wu et al. 2001).

The question we would like to address in this paper is: Can 
radiative cooling and nongravitational heating explain simultaneously 
the observed global X-ray properties of groups and clusters and 
the unresolved cosmic XRB ?  Previous studies have reached the
following conclusions: If one leaves the energy source problem aside, 
preheating model is in principle able to reproduce the observed 
$L_{\rm X}$-$T$ relation and excess entropy in the central regions of 
groups and clusters (e.g. Cavaliere, Menci \& Tozzi 1997, 1998; 
Balogh et al. 1999; Tozzi \& Norman 2001; Babul et al. 2002; etc.) 
and predict a consistent XRB with that observed (Pen 1999; 
Wu et al. 2001; Bryan \& Voit 2001).  As for 
radiative cooling, it has successfully accounted for the entropy 
floor seen in the central cores of groups and clusters 
(Voit \& Bryan 2001b). By properly truncating
the outer radii of X-ray surface brightness for groups and 
clusters in terms of current X-ray flux limits, 
one is also able to reproduce the observed steepening of the 
$L_{\rm X}$-$T$ relation of groups and clusters (Wu \& Xue 2002a; 
Voit et al. 2002).  In particular, the $L_{\rm X}$-$T$ relation 
found by hydrodynamical simulations in terms of cooling 
shows a good agreement with observations
(e.g. Muanwong et al. 2001, 2002; Borgani et al. 2002).
Consequently, it is crucial to reexamine whether or not  
cooling model is also sufficient enough to suppress the contribution of
the IGM to the soft XRB. If the negative result of  Wu et al. (2001) 
is confirmed,  it deserves to explore the combined effect of 
radiative cooling plus heating by supernovae 
on the X-ray properties of groups
and clusters and the soft XRB. Failure of all these efforts may
indicate that our current picture of the physical 
processes of the hot IGM in groups and clusters is incomplete
at some level.
Throughout this paper we assume a flat cosmological  
model ($\Lambda$CDM)
of $\Omega_{\rm M}=0.35$, $\Omega_{\Lambda}=0.65$ and $h=0.65$.

\section{Dark halos}

Dark matter distribution in a virialized halo is assumed to follow
the universal density profile suggested by
numerical simulations (Navarro, Frenk \& White 1997; NFW)
\begin{equation}
\rho_{\rm DM}(r)=\frac{\delta_{\rm ch}\rho_{\rm crit}}
                 {(r/r_{\rm s})(1+r/r_{\rm s})^2},              
\end{equation}
where $\delta_{\rm ch}$ and $r_{\rm s}$ are the
characteristic density and length of the halo, respectively,
and  $\rho_{\rm crit}$ is the critical density of the universe. 
In order to fix the two free parameters, $\delta_{\rm ch}$ and $r_{\rm s}$,
we first specify the concentration parameter $c=r_{\rm vir}/r_{\rm s}$ 
for a given halo of mass $M$ through 
the empirical fitting formula found by numerical simulations 
(Bullock et al. 2001) 
\begin{equation}
c=\frac{10}{1+z}\left(\frac{M}
  {2.1\times 10^{13}M_{\odot}}\right)^{-0.14}.
\end{equation}       
Next, we define the virial mass $M$ such that within the virial radius
$r_{\rm vir}$ the mean mass density of the dark halo is $\Delta_{\rm c}$ times 
the critical density of the universe:
\begin{equation}
M=\frac{4}{3}\pi r^3_{\rm vir}\Delta_{\rm c}\rho_{\rm crit},
\end{equation}
where for a flat, $\Lambda$CDM cosmological model, 
$\Delta_{\rm c}=18\pi^2+82[\Omega_{\rm M}(z)-1]-39[\Omega_{\rm M}(z)-1]^2$,
$\Omega_{\rm M}(z)=\Omega_{M}(1+z)^3/E^2$ and
$E^2=\Omega_{\rm M}(1+z)^3+\Omega_{\Lambda}$. 
Finally, we determine the virial 
temperature using cosmic virial theorem (Bryan and Norman 1998):
\begin{equation}
kT=1.39\;{\rm keV}\;
            \left(\frac{M}{10^{15}\;M_{\odot}}\right)^{2/3}
            \left(h^2 E^2\Delta_{\rm c}\right)^{1/3},  
\end{equation}
In which we have taken the normalization factor to be $f_T=1$.
We have also tested a lower value of $f_T=0.8$, and found that
our results remain almost unchanged.

We use the modified PS mass function by Sheth \& Tormen (1999) 
to describe the abundance and
evolution of virialized dark halos that grow from random-phase
Gaussian initial fluctuations:
\begin{equation}
dN=A\sqrt{\frac{2a}{\pi}}
       \left[1+\left(\frac{\delta_{\rm c}^2}
            {a\sigma^2}\right)^{p}\right]
    \frac{\bar{\rho}}{M}
    \frac{\delta_{\rm c}}{\sigma^2}   \frac{d\sigma}{dM}
    \exp\left(-\frac{a\delta_{\rm c}^2}{2\sigma^2}\right) dM,
\end{equation}
where $A=0.3222$, $a=0.707$, $p=0.3$, 
$\bar{\rho}$ is the mean cosmic density, $\delta_{\rm c}$ is the
linear over-density of perturbations that collapsed and virialized at
redshift $z$, $\sigma$ is the linear theory variance of the mass density
fluctuation in sphere of mass $M=4\pi\bar{\rho}R^3/3$.
We parameterize the power
spectrum of fluctuation $P(k)\propto k^nT^2(k)$ and take the fit
given by Bardeen et al. (1986) for the transfer function of
adiabatic CDM model $T(k)$. The primordial power spectrum is assumed
to be the Harrison-Zel'dovich case $n=1$. The mass variance for
a given $P(k)$ is simply
\begin{equation}
\sigma^2(M)=\frac{1}{2\pi^2}\int_0^{\infty}k^2P(k)W^2(kR)dk,
\end{equation}
where $W(x)=3(\sin x-x\cos x)/x^3$ is the Fourier
representation of the window function. The amplitude in the power
spectrum is determined using the rms fluctuation
on an $8$ $h^{-1}$ Mpc scale, $\sigma_8$. 
We adopt a normalization parameter of $\sigma_8=0.9$ suggested 
by weak gravitational lensing measurements 
(see Refregier, Rhodes \& Grith 2002 for a recent summary), and 
then demonstrate the influence of
$\sigma_8$ on the evaluation of XRB using a lower value of 
$\sigma_8=0.7$.

\section{IGM}

\subsection{Global quantities}

Following conventional definition, we use
\begin{equation}
S=\frac{kT}{n_{\rm e}^{2/3}}
\end{equation}
to measure the entropy of IGM, in which $T$ and $n_{\rm e}$ are
the electron temperature and number density, respectively.
The total X-ray luminosity in terms of thermal bremsstrahlung is
\begin{equation}
L_{\rm X}=\int n_{\rm e}n_{\rm H}\Lambda(T)dV,
\end{equation}       
where $n_{\rm H}$ is the number density of hydrogen, and 
$\Lambda(T)$ is the cooling function which is calculated from  
the Raymond-Smith (1977) code. Whenever theoretical predictions
are compared with observations, we work with the emission-weighted 
temperature
\begin{equation}
T=\frac{\int T(r)\Lambda(T)n_{\rm e}n_{\rm H}dV}
           {\int \Lambda(T)n_{\rm e}n_{\rm H}dV}.
\end{equation} 
Furthermore, we assume that the IGM with and without 
cooling/preheating is always in hydrostatic equilibrium with the 
underlying gravitational potential of groups/clusters dominated
by dark matter:
\begin{equation}
\frac{1}{\mu m_{\rm p} n_{\rm e}(r)}\frac{d[n_{\rm e}(r)kT(r)]}{dr}=
  -\frac{GM_{\rm DM}(r)}{r^2},
\end{equation}   
where $\mu\approx 0.593$ is the mean molecular weight. 
Finally, we evaluate the total XRB intensity at frequency $\nu$ by
integrating the X-ray emission of all halos 
over mass range and redshift space 
\begin{equation}
J(\nu)=\int\int\frac{dL_{\rm X}/d(h_{\rm p}\nu)}{4\pi D_{\rm L}^2(z)}
        \frac{dN}{dVdM}\frac{dV}{d\Omega dz}dMdz,
\end{equation}
where $h_{\rm p}$ is the Planck constant, $D_{\rm L}$ is the luminosity
distance, and $dN/dVdM$ is the mass function given by equation (5).

\subsection{No cooling and no heating}

In the absence of radiative cooling and nongravitational heating, 
we assume that IGM traces dark matter
\begin{equation}
\rho_{\rm gas}(r)=f_{\rm b}\rho_{\rm DM}(r).
\end{equation}
We solve the equation of hydrostatic equilibrium under the boundary
restriction $T(\infty)=0$, which yields
\begin{equation}
kT(r)=kT^*\frac{r}{r_{\rm s}}\left(1+\frac{r}{r_{\rm s}}\right)^2
       \int_{\frac{r}{r_{\rm s}}}^{\infty}
       \frac{(1+x)\ln (1+x)-x}{x^3(1+x)^3}dx,
\end{equation} 
where $kT^*=4\pi G \mu m_{\rm p}\delta_{\rm ch}\rho_{\rm crit}r_{\rm s}^2$.
This allows us to calculate straightforwardly the X-ray luminosity, 
entropy and total XRB. 
Our numerical computations show that these quantities
are rather insensitive to the radial variation of temperature described
by equation (13).  Consequently, we may adopt a constant temperature 
instead of equation (13) in the evaluation of the global X-ray 
properties of groups and clusters and the XRB. 
It is well known that the results predicted by 
this so-called `self-similar' model (hereafter Model I)  depart 
remarkably from X-ray observations. 
Here we use this model as a reference point only.

\subsection{Radiative cooling}

Conservation of energy ensures that the energy loss due to 
bremsstrahlung emission is balanced by the decrease in the specific 
energy of IGM in groups/clusters:
\begin{equation}
\frac{3}{2}n_{\rm t}kT=n_{\rm e}n_{\rm H}\Lambda(T)t_{\rm c},
\end{equation}
where $n_{\rm t}$ is the total number density of the IGM. 
Setting the cooling time $t_{\rm c}$ to equal the age of groups/clusters,
or simply the age of the universe at the redshift of the groups/clusters
determines the cooling radius $r_{\rm cool}$ and the total mass
$M_{\rm cool}$ of the cooled material if the IGM is assumed
to follow the dark matter distribution before cooling. 
Following Voit \& Bryan (2001b) and Wu \& Xue (2002a), we can find 
the distribution of the remaining IGM after cooling by solving the
equation of hydrostatic equilibrium under the conservation of 
total baryonic mass
\begin{equation}
M_{\rm cool}(r_{\rm vir})+M_{\rm gas}(r_{\rm vir})=
             f_{\rm b}M, 
\end{equation}  
and the conservation of entropy 
\begin{equation}
\frac{T(r)}{[n_{\rm e}(r)]^{2/3}}=
         \frac{T^0(\bar{r})}{[n_{\rm e}^0(\bar{r})]^{2/3}},
\end{equation}
where $r$ is related to $\bar{r}$ through 
\begin{equation}
M_{\rm gas}(r)=M^0_{\rm gas}(\bar{r})-M_{\rm cool}.
\end{equation}
Here we use superscript $^0$ to denote the quantities
before cooling. The above two equations demonstrate that  
the IGM originally distributed between $r_{\rm cool}$ and 
$\bar{r}$ before cooling is transported to a smaller region 
of radius $r$ after cooling. 
We consider an evolving metallicity model of $Z=0.3Z_{\odot}(t/t_0)$ 
(Model II) and a constant metallicity model of  $Z=0.3Z_{\odot}$
(Model III), where $t_0$ is the present age of the universe.

\subsection{Preheating}

A phenomenological treatment of (pre)heating is to raise 
the specific energy or entropy of the IGM in groups and clusters 
to a certain level 
regardless of whatever the energy sources would be. Usually, an
energy budget of 0.1 -- 3 keV/particle or an entropy of 
50 -- 400 keV cm$^2$, depending on the epoch and
environment of (pre)heating, is needed in order to reproduce
the observed X-ray properties of groups and clusters.
Here we consider two simple approaches to demonstrating the 
effect of (pre)heating on the X-ray properties of groups and clusters.

Model IV: We begin with the IGM distribution predicted by self-similar model 
(Model I), and then simply raise the specific energy of each particle 
in groups and clusters by a constant amount of $k\Delta T$:
\begin{equation}
kT(r)=k\Delta T+kT_{\rm self}(r).
\end{equation}
The corresponding distribution of electron number density $n_{\rm e}(r)$ 
can be obtained by solving the equation of hydrostatic equilibrium. 
A critical point in such an exercise is the boundary condition. To a
first approximation we may assume that the gas density at virial
radius $r_{\rm vir}$ is universal, $n_{\rm e}(r_{\rm vir})=
(f_{\rm b}/\mu_{\rm e}m_{\rm p})\rho_{\rm DM}(r_{\rm vir})$. 
Nevertheless, one should keep in mind that this restriction may fail 
for low-mass systems. Another commonly adopted, inaccurate method 
is to take the total gas mass fraction within $r_{\rm vir}$ 
to be universal 
$M_{\rm gas}(r_{\rm vir})=f_{\rm b}M$.  
Actually, one still has no better choice
of boundary condition in the determination of the gas distribution 
from the equation of hydrostatic equilibrium.

Model V: Instead of raising the specific energy of the IGM particles,
we add a constant entropy floor $\Delta S$ to the entropy profile 
given by self-similar model $S_{\rm self}(r)$
\begin{equation}
S(r)=\Delta S+S_{\rm self}(r).
\end{equation}
We then solve the equation of hydrostatic equilibrium to get the
electron pressure distribution $P_{\rm e}=n_{\rm e}kT$
\begin{equation}
P_{\rm e}(r)=\left[\frac{2G\mu m_{\rm p}}{5}
                  \int_r^{\infty} \frac{M_{\rm DM}(r)}{S^{3/5}(r)r^2}dr
	     \right]^{5/2},
\end{equation}
where we have adopted the boundary condition $P_{\rm e}(\infty)=0$.
Note that this restriction should not be taken too literally 
because the equation of hydrostatic equilibrium may break 
down beyond $r_{\rm vir}$.

\subsection{Radiative cooling plus heating}

We first work with the IGM temperature and entropy distributions,
$T_{\rm cool}(r)$ and $S_{\rm cool}(r)$, 
predicted by cooling (Model II), 
and then raise either the specific energy of each particle by a constant 
amount of $k\Delta T$ (Model IV) or the entropy by a constant 
floor of  $\Delta S$ (Model V)
so that the new temperature (Model VI) and entropy profiles (Model VII) 
become  
\begin{equation}
T(r)=\Delta T+T_{\rm cool}(r),
\end{equation}
and
\begin{equation}
S(r)=\Delta S+S_{\rm cool}(r),
\end{equation}
respectively.
In a way similar to Models IV and V, we can find the new density profile of
the IGM for our combined model of radiative cooling plus heating. 
Note that our heating model differs from the conventional
preheating model. For the latter, the IGM is heated before 
cooling, while in our model we address the issue of 
how large an extra energy budget should still be added to the IGM 
after the cooled gas is removed from groups and clusters.

The parameters of the seven IGM models are summarized in Table 1.                

 \begin{table*}
 \vskip 0.2truein
\caption{Summary of the parameters and results for the IGM models.}
 \vskip 0.2truein
 \begin{tabular}{cccccccc}
 \tableline
 \tableline
Model   &  cooling & heating &  metallicity ($Z_{\odot}$)  & 
           $S$-$T$  &   $L_{\rm X}$-$T$  &  XRB ($\sigma_8=0.9$)   
           & XRB ($\sigma_8=0.7$) \\
 \tableline
I       &    no    &  no   &  $0.3(t/t_0)$ & fail & fail & fail & fail \\
II      &    yes   &  no   &  $0.3(t/t_0)$ & ok & fail/ok$^*$ & fail & fail \\
III     &    yes   &  no   &  $0.3$  &       ok & fail/ok$^*$ & fail & fail \\
IV      &    no    &  yes($\Delta T$)  &  $0.3(t/t_0)$ & ok & ok  & fail& ok \\
V       &    no    &  yes($\Delta S$)  &  $0.3(t/t_0)$   & ok & ok & fail& ok\\
VI      &    yes   &  yes($\Delta T$)  &  $0.3(t/t_0)$   & ok & ok & fail&ok \\
VII     &    yes   &  yes($\Delta S$)  &  $0.3(t/t_0)$   & ok & ok & fail&ok \\
 \tableline
 \end{tabular}
 \parbox {8.5in}{$^*$Acceptable if a truncated radius is properly 
                 introduced in terms of X-ray surface brightness limit.}  
 \end{table*}

\section{Results}

Our strategy is as follows: For each model listed in Table 1,
we first compute the expected entropy and X-ray luminosity 
distributions of groups and clusters, and then compare with X-ray 
measurements, which constitutes a fundamental test for 
each of the proposed models. 
For the heating models (Model IV and Model V) and cooling plus heating 
models (Model VI and Model VII), this allows us
to work out the most probable energy budget $k\Delta T$ or $\Delta S$  
with which the observed entropy and X-ray luminosity distributions 
of groups and clusters can be simultaneously accounted for.
We then compute the XRB
predicted by these models, and examine whether their 
predictions are compatible with current observational limits.

\subsection{Entropy distribution}

The entropy at $0.1r_{\rm vir}$ against the emission
weighted temperature $T$ predicted by all the models   
is shown in Figure 1, together with the updated measurements of 
$S(0.1r_{\rm vir})$ by Ponman et al. (1999), Lloyd-Davies et al. (2000) 
and Xu, Jin \& Wu (2001). 
Essentially, all models except self-similar model
can roughly reproduce the overall $S(0.1r_{\rm vir})$-$T$ distributions. 
In the preheating model (Fig.1b and Fig.1c) and the combined model (Fig.1d) 
we have tested a set of
energy input values $k\Delta T$ and $\Delta S$, in attempt to  
estimate the most probable parameters which give the closest fits of   
both the $S$-$T$ distribution and the $L_{\rm X}$-$T$ relation (see below).
In several cases we have also shown the curves for two sets of 
$k\Delta T$ and $\Delta S$ around the best-fit values in order
to demonstrate how sensitively the results depend on 
the input energy/entropy.   
While there are some differences in the predicted  $S$-$T$ relation
among various IGM models 
and for different input $k\Delta T$ or $\Delta S$ values
especially below temperature $T\sim1$ keV,
a decisive conclusion regarding which model provides
the most satisfactory explanation of the entropy data
cannot be drawn, because of sparse data points
and associated large uncertainties.
It is emphasized that the entropy profile expected from
radiative cooling alone (Fig.1a) is consistent with observed data
(Voit \& Bryan 2001b), although with an additional energy 
supply by preheating model the predicted entropy distribution 
seems to yield a better fit to the data points for low-temperature 
systems of $T<1$ keV (Fig.1d).

\subsection{$L_{\rm X}$-$T$ relation}

We now compare the bolometric X-ray luminosity -
temperature relations of groups and clusters predicted by
different IGM models with X-ray observations. For the latter  
we use the catalog of X-ray groups and clusters compiled by 
Wu et al. (1999) and Xue \& Wu (2000). The updated sample 
contains 57 groups and 192 clusters whose X-ray temperature
and luminosity are both available. We have converted the X-ray
luminosity in the Einstein-de Sitter universe 
into the one in the $\Lambda$CDM cosmological model. 

The $L_{\rm X}$-$T$ relations predicted by cooling models II and III
are shown in Figure 2.
While there is a good agreement between the predicted and 
observed  $L_{\rm X}$-$T$ relations on cluster scale  
$T>4$ keV, cooling models become to be insufficient to 
recover the observed data at temperature below $\sim4$ keV. 
One possible reason for this discrepancy may arise purely 
from selection effect: The theoretically predicted X-ray luminosity 
accounts for all emission inside virial radii of 
groups, while current X-ray observations have not corrected
for lost flux falling out the detection aperture for majority
of groups with $\beta$ parameters less than $1/2$
(e.g. Ponman et al. 1996; Helsdon \& Ponman 2000). 
For this reason, following Wu \& Xue (2002a) and Voit et al. (2002)
we re-calculate the X-ray luminosity for Model II by excluding 
the contributions of the IGM in the outer regions of groups and clusters 
set by a given X-ray surface brightness limit $S_{\rm limit}$. 
The expected X-ray luminosity of clusters is almost unaffected by this 
truncation. However, the X-ray luminosity of groups suffers 
seriously from this selection effect, and the resulting  
$L_{\rm X}$ drops remarkably at low-temperature end.
This may partially eliminate our concern about the efficiency of
cooling in the explanation of $L_{\rm X}$-$T$ relation.  
Moreover, the $L_{\rm X}$-$T$ relations of groups and clusters with and 
without the inclusion of the cosmic variation of metallicity 
(II and III) show little evolution at least out to $z=1$. 

Figure 3 illustrates the $L_{\rm X}$-$T$ relations predicted by 
preheating model (IV and V). Adding an extra energy of 
$k\Delta T=0.25 $ keV to each particle in groups and clusters 
roughly reproduces the observed $L_{\rm X}$-$T$ relation. 
Note, however, that there is a small but significant vertical 
shift between our predicted and observed data in the temperature
range $2<T<8$ keV (Fig.3a). We have then tried a larger value of 
$k\Delta T=0.3$ keV, which indeed reduces somewhat the difference. 
A further increase of  $k\Delta T$ leads the entropy curve 
in Fig.1b to rise too high to be consistent with observations. 
When $k\Delta T$ exceeds the virial temperature of a given group, we 
assume that the IGM cannot be trapped in the system. This
is responsible for the cutoff of our $L_{\rm X}$-$T$ relation at low
temperature end.  
For Model V, in which we have raised the entropy by a
constant floor of $\Delta S=120$ keV cm$^2$, 
the predicted $L_{\rm X}$-$T$ relation of clusters matches perfectly 
the observed one. 
For low-temperature groups, the agreement
becomes only marginal. Increasing slightly the value of 
$\Delta S$ improves the match. But, it meanwhile raises
the entropy distribution (Fig.1c). We have tried a value 
of $\Delta S=200$ keV cm$^2$, and found that the entropy
at low-temperature becomes too large to be reconciled with
the observed data. In a word, it turns out that Model V is marginally 
acceptable. Alternatively, our numerical computations show 
that there is no apparent cosmic evolution 
of the predicted $L_{\rm X}$-$T$ relations for both Model
IV and Model V, regardless of whether or not the metallicity
varies with cosmic time in terms of  $0.3Z_{\odot}(t/t_0)$.

Because either cooling or preheating alone can reproduce or 
marginally reproduce the observed $L_{\rm X}$-$T$ and 
$S(0.1r_{\rm vir})$-$T$ relations, it is naturally expected that a 
combination of the two mechanisms should be successful in  
the explanation of the two relationships. 
Indeed, within the framework of radiative cooling, an additional
energy input of $k\Delta T=0.1$ keV in Model IV and an 
entropy input of $\Delta S=50$ keV cm$^2$ in Model IIV 
both result in the X-ray luminosity and entropy distributions
which agree nicely  with X-ray observations (see Figure 4).
We have also considered the situation of a constant metallicity
of $0.3Z_{\odot}$ and at high redshifts out to $z=1$, 
and found that the corresponding modifications are only minor.

\subsection{XRB}

A considerably large fraction of soft and hard XRB has been 
resolved into discrete sources (e.g. 
McHardy et al. 1998; Hasinger et al. 1998, 2001;
Mushotzky et al. 2000; Giacconi et al. 2001, 2002; Tozzi et al. 2001;
Hornschemeier et al. 2001; Rosati et al. 2002; Bauer et al. 2002; etc.).
The maximum admitted ranges of the unresolved flux at different energy
bands have been summarized in Wu \& Xue (2001). Here we only add a new 
upper limit in the 2-8 keV band from the 1 Ms Chandra observation of 
the Chandra Deep Field North (Cowie et al. 2002):
$0.5 \times10^{-11}$ ergs s$^{-1}$ cm$^{-2}$ deg$^{-2}$.
It should be kept in mind that the diffuse X-ray emission from some of
the nearby, bright galaxies, groups and clusters has also been included 
in the current resolved soft XRB, although the fractions of these 
diffuse X-ray sources in the resolved and unresolved XRB 
are still uncertain. In this regard, the residual soft XRB after 
the removal of the discrete sources may not be taken to be a 
very stringent upper limit on the contributions of 
groups and clusters.

The expected XRB spectra from different IGM models  
in terms of Equation (11) are shown and 
compared with the observational upper limits in Figure 5. 
Meanwhile, we also demonstrate the XRB by summing up
the contributions of groups and clusters described by their
X-ray luminosity functions (Wu \& Xue 2001). The good agreement 
between the XRB produced by the `known' population of groups and 
clusters and the current limits indicates that
the diffuse IGM confined in groups and clusters is probably the
major source of the unresolved soft XRB. This may help to eliminate
the above concern that some of the bright groups and clusters 
have been resolved and thus removed from the residual XRB, and 
the unresolved XRB may not constitute a robust constraint on 
the diffuse IGM of groups and clusters. 
Actually, nearby bright and massive clusters (e.g. $z<0.2$ and 
$M>5\times10^{14}M_{\odot}$)  only make a minor  
contribution to the total soft XRB (Wu \& Xue 2001; see also 
Figure 5).

It turns out from Figure 5 that the soft XRB predicted by the 
self-similar model (Model I) vastly exceeds the observational
limits (Pen 1999; Wu et al. 2001). 
The huge difference of up to two orders of magnitude
at $E\sim0.1$ keV implies that the IGM should have 
a much shallower distribution than dark matter 
especially in low-mass systems. Preheating was thus  
advocated as a potentially important mechanism    
to break the similarity between dark matter and IGM.

We first begin with the radiative cooling scenario.
The predicted XRB remains almost unchanged if a constant metallicity
of $0.3Z_{\odot}$ (Model III) is replaced by a time-varying quantity of 
$0.3Z_{\odot}(t/t_0)$ (Model II). Actually, our numerical computations 
show that this conclusion applies to all the models in Table 1.
In the hard energy band $E>2$ keV, the XRB produced by groups and 
clusters within the framework of cooling is well below the observational
limits. By contrast, in the soft energy band $E\approx0.1$-$2$ keV
the cooling results are about 2--4 times larger than the current 
upper limits placed on contribution from diffuse IGM to the XRB.
This is consistent with previous findings by Wu et al. (2001)
based on N-body simulations of halo merger trees coupled with 
semi-analytic models.  We have also studied the contributions 
of different halos to the total XRB, and found that 
most of the soft XRB is produced by groups 
of mass $M\sim10^{14}$ $M_{\odot}$. Note that 
very low-mass halos make almost no contribution to the XRB
because they contain very little hot IGM due to their 
too short cooling time (see Fig.5a). It appears that although cooling provides 
a more or less reasonable explanation of the X-ray luminosity and
entropy distributions, it is insufficient (by a factor of 
2--4) to account for the unresolved soft XRB.

We now turn to preheating model. For Model IV, in which
the specific energy of each particle in groups and
clusters is raised by a constant amount
of $k\Delta T=0.25 $ keV,  we have successfully recovered 
the observed $S(0.1r_{\rm vir})$-$T$ relation (see Fig.1b) and roughly
reproduced the $L_{\rm X}$-$T$ relation (see Fig.3). 
Our expected XRB from this model is shown in Fig.5b.
Unfortunately, we reach a result very similar to the cooling
prediction discussed above: Model IV fails to efficiently 
suppress the X-ray emission of groups and clusters to a level 
below the unresolved XRB,
and the difference in the soft energy band $E=0.1$-$2$ keV 
is approximately by a factor of  2--4.

For another preheating model V, in which we have raised the
entropy of IGM by a constant floor $\Delta S$ instead of
$k\Delta T$ for Model IV, we present the XRB in Figure 5c
for two choices of $\Delta S$, 120 and 200 keV cm$^2$,
respectively. The former nicely reproduces the entropy
distribution but only marginally explains the $L_{\rm X}$-$T$ relation,
while the latter turns to be successful in the recovery of 
the $L_{\rm X}$-$T$ relation but results in an overestimate of 
the entropy distribution. Now, in terms of their predicted
XRB spectra alone, the model with $\Delta S=200$ keV cm$^2$ 
becomes acceptable, while the entropy level of  
$\Delta S=120$ keV cm$^2$ is not sufficiently large
to reduce the XRB to the observational limits. 
This is roughly consistent with simulation results that 
an entropy floor of $100-200$ keV cm$^2$ is needed  
to suppress the expected unresolved X-ray background 
below the observational limits (Voit \& Bryan 2001a).

Finally, we come to the cooling plus heating model.
Adding an energy budget of $k\Delta T=0.1$ keV to each particle
in addition to radiative cooling (Model VI) has allowed us
to reproduce remarkably well the entropy and X-ray luminosity 
distributions of groups and clusters, as are shown in Fig.1d
and Fig.4.  
The XRB predicted from this model is consistent with the 
observational constraints above $T\sim 1$ keV (see Fig.5d).
Nonetheless, the theoretical prediction becomes to be 
larger than the observational limits by a factor of 2--3 
in the energy range from 0.1 to 1 keV. Because of the apparent 
success of Model VI in the explanation of the X-ray
properties of groups and clusters, it deserves further investigation
of whether this disagreement arises from other effects.
First, we exclude the contribution of  nearby ($z<0.2$) and
massive ($M\geq5\times10^{14}M_{\odot}$) clusters to the
XRB, in the sense that these bright, diffuse X-ray sources may 
have been resolved by current deep X-ray observations. 
This leads to a moderate decrease of the expected XRB in
high energy band but has only a minor effect on the soft XRB.
Second, we attempt to include the absorption of X-ray emission  
by the neutral hydrogen in groups and clusters 
as a consequence of radiative cooling. To this end, 
we assume a simple King model for the density distribution
of the cooled material, in which we take the core size to be
$r_{\rm c}=0.1r_{\rm vir}$. Moreover, we normalize this density profile
using the total mass of the cooled IGM given by cooling 
scenario, Equation (14).
The total X-ray luminosity at frequency $\nu$ 
with self-absorption for a given halo reads
\begin{eqnarray}
\frac{dL_{\rm X}}{dh_{\rm p}\nu}&=&1.15\times10^{56}
         \left(\frac{{\rm keV}\;{\rm s}^{-1}}{{\rm keV}}\right)
	 \left(\frac{r_{\rm vir}}{\rm Mpc}\right)^3
         \int_0^{\pi} \sin\theta d\theta \nonumber \\
      & &\int_0^c e^{-\sigma N_i}
         \left(\frac{n_{\rm H}}{n_{\rm e}}\right)
	 \left(\frac{d\Lambda/dh_{\rm p}\nu}
	 {10^{-23}{\rm ergs}\;{\rm s}^{-1}\;{\rm cm}^3\;{\rm keV}^{-1}}\right)
	 \nonumber\\
      & &\left(\frac{n_{\rm e}}{10^{-2}{\rm cm}^{-3}}\right)^2
	 x^2dx,
\end{eqnarray}
where $\sigma$ is the effective absorption cross-section, 
$N_i$ is the hydrogen column density, and
\begin{eqnarray}
\sigma N_i&=&0.31
	      \left(\frac{\sigma(h_{\rm p}\nu)}{10^{-22}{\rm cm}^2}\right)
              \left(\frac{n_{\rm H0}}{10^{-3}{\rm cm}^{-3}}\right)
	      \left(\frac{r_{\rm c}^3}{r_{\rm c}^2+r^2\sin^2\theta}\right)
	      \nonumber\\
          & & \left(\frac{r\cos\theta}{\sqrt{r^2+r_{\rm c}^2}} 
                   +\frac{\sqrt{r_{\rm vir}^2-r^2\sin^2\theta}}
                         {\sqrt{r_{\rm vir}^2+r_{\rm c}^2}} \right),
\end{eqnarray}
in which $n_{\rm H0}$ is the central number density of neutral hydrogen,
$r$ and $r_{\rm c}$ are in units of Mpc. We adopt the fitting
formula of Morrison \& McCammon (1983) to calculate $\sigma$. 
The XRB with the correction for hydrogen absorption is shown in Figure 5d. 
It appears that the inclusion of self-absorption results in
a significant decrease of the XRB below $E\sim0.2$ keV, while 
the rest XRB at $E>0.2$ keV remains almost unchanged.

As for Model VII, the fact that the $L_{\rm X}$-$T$ relation at the 
low temperature slightly exceeds the observed data (see Fig.3),
despite the excellent agreement between the model-predicted
entropy distribution for $\Delta S=50$--$100$ keV cm$^2$ 
and the observations, indicates that the expected XRB at soft energy 
band will exhibit an excess relative to the observational limits.
Indeed, although raising the extra entropy floor  to 
$\Delta S=100$ keV cm$^2$ leads to a decrease of the expected
XRB relative to the cooling model prediction, in the low energy range 
$E<1$ keV the prediction by model VII  
and observation limits still differ by a factor of $\sim2$.

We notice, however, that the XRB spectra predicted by cooling,
preheating and cooling plus heating models
actually have similar shape except that they are displaced upward 
in amplitude by a factor of 2--5 as compared with the unresolved soft
XRB. This may suggest a common origin of 
the discrepancies if they are not due to
our incomplete knowledge of gas physics in groups and clusters.
We thus recalculate the XRB for all the models adopting a lower value of
the normalization of the mass function of clusters and groups
$\sigma_8=0.7$, instead of $\sigma_8=0.9$, 
(e.g. Seljak 2001; Schuecker et al. 2002; etc.). 
The corresponding XRB spectra are illustrated in Figure 6. 
It appears that cooling model is still insufficient to bring
the expected XRB to the observational limits in the soft energy band
below $E<1$ keV. However, both preheating (especially Model V) and 
cooling plus heating models now yield the XRB spectra which agree nicely
with the unresolved XRB. Perhaps, this indicates that the diffuse XRB 
could also be used as an independent constraint on the normalization 
of group and cluster abundance.

\section{Conclusions}

We have explored a set of simple analytic models for the distribution and
evolution of the IGM in groups and clusters, aiming at highlighting 
the dominant physical process for the hot IGM,   which may
complement our understanding of the essential physics in addition to
employment of hydrodynamical simulations.
In particular, we have addressed the question of whether the prevailing
scenarios, namely, radiative cooling, preheating and a combination 
of cooling and heating, can explain both the global observed X-ray 
properties (entropy distribution and $L_{\rm X}$-$T$ relation) 
of groups and clusters and the observational limits on the 
contribution of the diffuse IGM in virialized halos to the XRB
within the framework of standard CDM structure formation 
with an amplitude of matter power spectrum $\sigma_8=0.9$.
Our main conclusions are summarized as follows:

Without radiative cooling and extra heating in addition to 
gravitational shocks and adiabatic compression, the IGM appears to
be too concentrated in halos to explain all the X-ray observations,
especially in low-mass systems.

Inclusion of radiative cooling may allow one to marginally reproduce the 
entropy floor seen in the central regions of groups and clusters
(Voit \& Bryan 2001b). However, it is still insufficient to  account for the
steepening of the $L_{\rm X}$-$T$ relation on group scale,
if X-ray surface-brightness bias is not included, and the observational
upper limits on the diffuse XRB. For the latter, the difference 
is by a factor of $2$--$5$, consistent with the previous findings
of Wu et al. (2001). Although cooling is certainly an important process 
in the formation and evolution of galaxies, groups and clusters, and
also for the explanation of the entropy floors shown in Fig.1, and even
responsible for the scale-dependence of the IGM mass fraction 
(Wu \& Xue 2002b), energy feedback from galaxy formation should be
included in order to resolve the inefficiency problem and the cooling
crisis (Balogh et al. 2001).

A phenomenological treatment of preheating, regardless of whatever 
the heating sources and mechanisms would be, may allow us to both reproduce 
the X-ray luminosity distribution of groups and 
clusters and explain the observational limits on the XRB from IGM, 
provided that the level of entropy floor can be raised 
to as large as 200 keV cm$^2$ at present epoch. 
Apparently,  the entropy profile in this case exceeds the 
measurements (Fig.1c). When we fix the value of input entropy floor to 
$\Delta S=120$ keV cm$^2$, which gives a nice fit to the observed
central entropy distribution of groups and clusters, our predicted 
$L_{\rm X}$-$T$ relation and XRB both show an excess at 
low-temperature/energy range. In a word, a naive preheating model
may meet difficulty in the recovery of all the observational 
phenomena. 

Our cooling model with extra heating also fails when the predicted soft 
XRB is compared with the observational limits. It has been shown that
the inclusion of self-absorption by the cold gas in groups and 
clusters and the exclusion of the contribution of nearby, massive clusters
to the XRB do not alter the conclusion significantly. 
Of course, our cooling plus
heating scenario differs from the commonly used preheating plus 
cooling model in which the IGM was preheated before radiative cooling
comes into effect.  

Despite of their oversimplification,
our analyses as a whole suggest that none of the present 
models can simultaneously account for the observed
X-ray properties of groups and clusters and the residual soft XRB.
Actually, the latter constitutes a very stringent constraint on
the content and physical processes of the IGM in groups and clusters.
Recall that the current observational limits even do not account for 
the fact that a large fraction of the 
unresolved soft XRB ($\sim10\%$ of the total XRB) which we have adopted 
in this paper may still arise 
from the X-ray emission of discrete sources (e.g. faint galaxies).  
More sophisticated models in combination with hydrodynamical simulations
will thus be needed to further address the issue.

Finally, the theoretically predicted XRB depends sensitively on 
the normalization of group and cluster abundance. If a lower value of 
$\sigma_8=0.7$ is adopted instead of $\sigma_8=0.9$, it has been shown
that the expected XRB spectra from preheating model and 
cooling plus heating model agree nicely with current observational limits. 
In other words, the observational constraints on the unresolved
diffuse XRB lends support to a lower amplitude of matter 
power spectrum $\sigma_8\approx0.7$ 
(e.g. Seljak 2001; Schuecker et al. 2002). 
Indeed, the discrepancy between theoretical predictions 
(at least for preheating model and cooling plus heating model) and 
X-ray observations, if it is not a result of the oversimplification of 
our analytic models, may have simply arisen from 
the too high normalization of the present matter power spectrum.

\acknowledgments
We thank an anonymous referee for valuable suggestions.
This work was supported by the National Science Foundation of China, 
and the Ministry of Science and Technology of China, under Grant
No. NKBRSF G19990754.

\clearpage

\begin{figure}
\epsscale{0.8}
\plotone{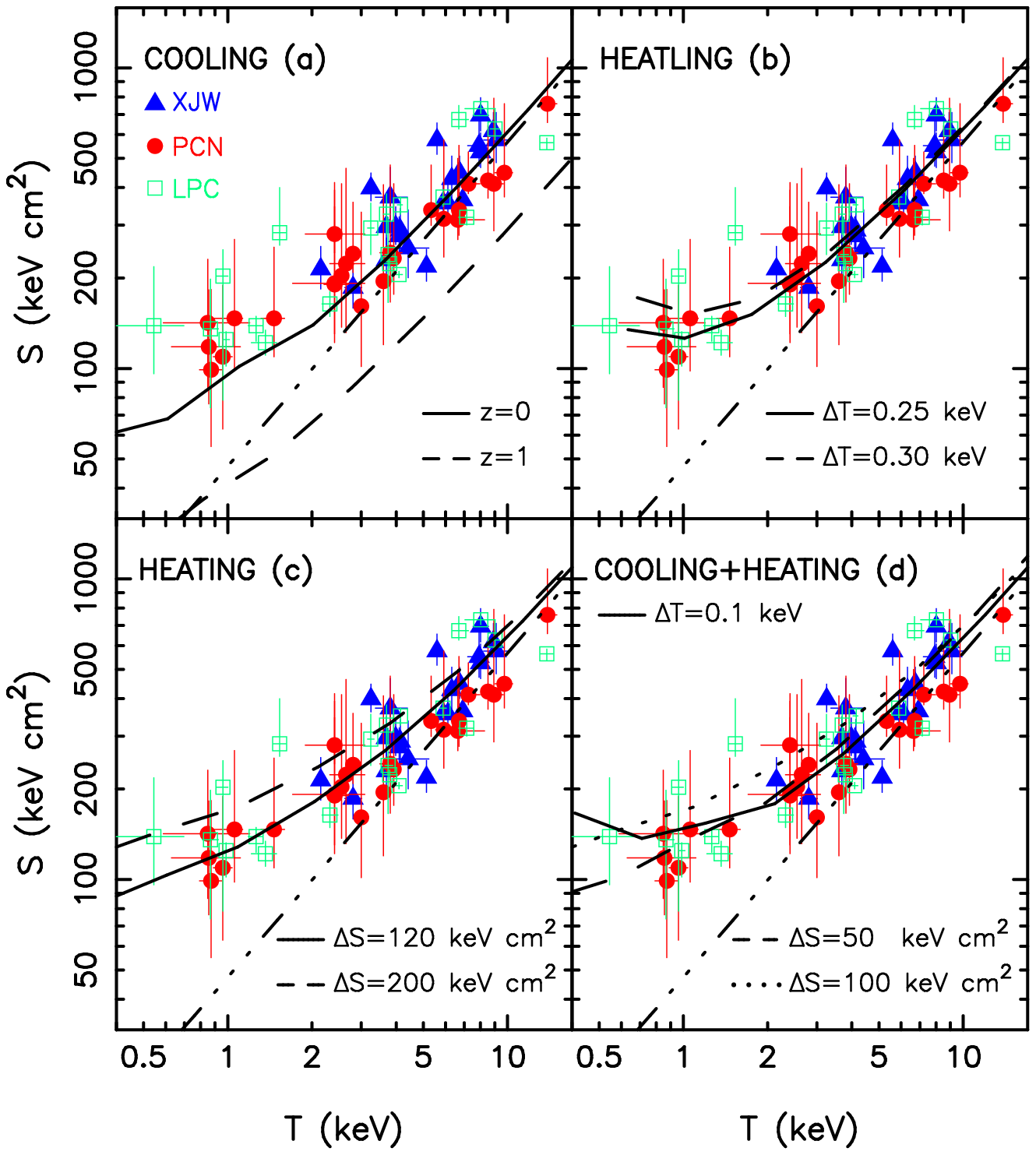}
\caption{Entropy distributions of groups and clusters measured at
$0.1r_{\rm vir}$. Observational data are taken from 
Ponman et al. (1999; PCN), Lloyd-Davies et al. (2000; LPC)
and Xu et al. (2001; XJW). Dash-dot-dot-dot line is the self-similar
model (Model I) prediction. For cooling model (a) 
the entropy profile at $z=1$ is also illustrated. 
\label{fig1}}
\end{figure}

\clearpage 

\begin{figure}
\epsscale{0.5}
\plotone{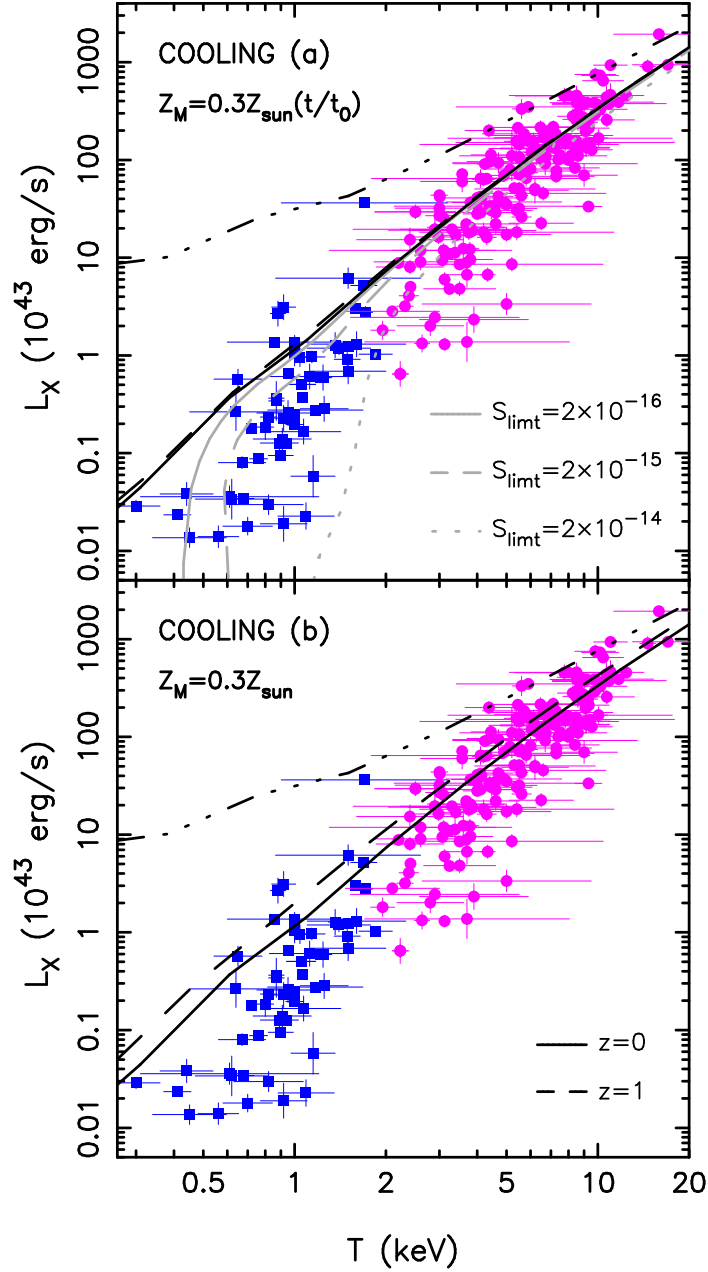}
\caption{The X-ray luminosity - temperature relations predicted by
cooling models II and III are compared with observations, in which
we consider a time-evolving metallicity (upper panel) and a 
constant  metallicity of $0.3Z_{\odot}$ (lower panel).
Meanwhile, in the upper panel we demonstrate how the $L_{\rm X}$-$T$ relation
is modified by different X-ray flux limits $S_{\rm limit}$ in units of 
ergs s$^{-1}$ arcmin$^{-2}$ cm$^{-2}$ (grey lines). 
The $L_{\rm X}$-$T$ relations at redshift $z=1$ for the two models
are also illustrated (dashed lines). For comparison, the results for 
self-similar model are shown by the dash-dot-dot-dot lines.
\label{fig2}}
\end{figure}

\clearpage 

\begin{figure}
\epsscale{0.5}
\plotone{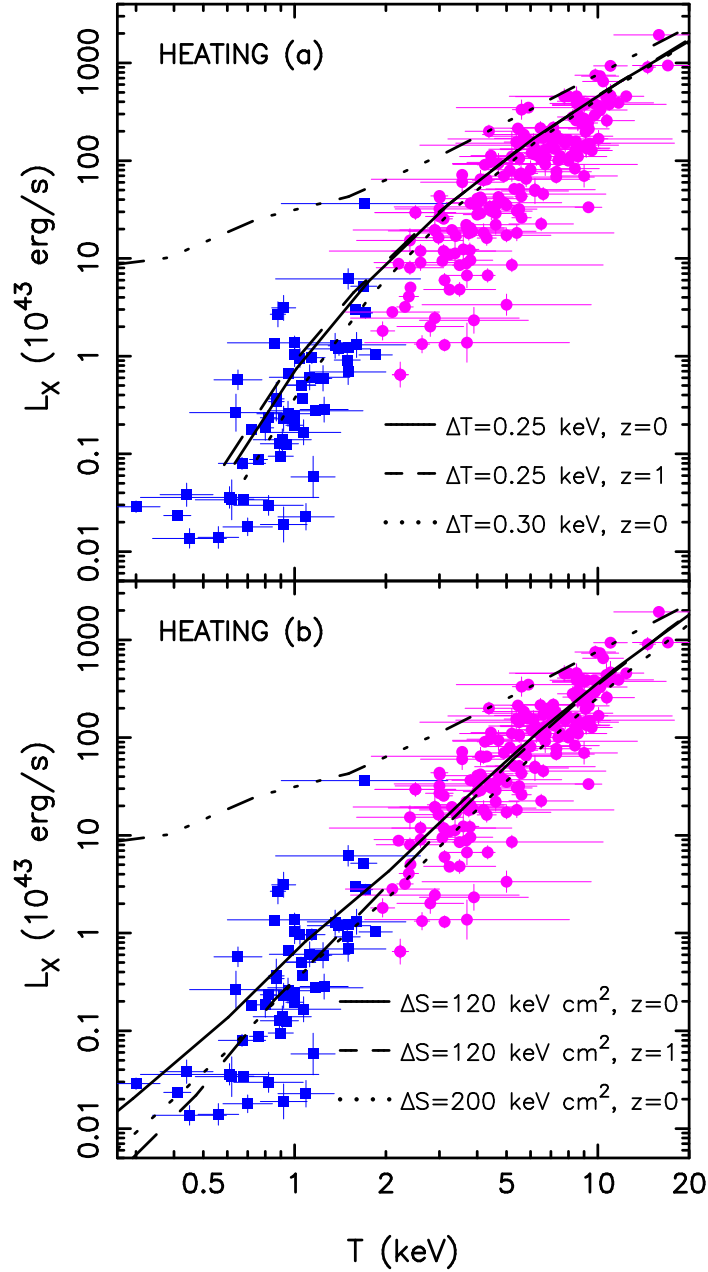}
\caption{The X-ray luminosity - temperature relations predicted by
preheating models IV (upper panel) and V (lower panel). For Model IV 
we show the results for two energy inputs: $k\Delta T=0.25$ and 0.3
keV, respectively.
Note that there is a slight excess of the predicted 
$L_{\rm X}$ relative to the observed data
in the temperature range from $\sim2$ to $\sim8$ keV.
For Model V two different constant 
entropy floors are added to the IGM: $\Delta S=120$ and 200 keV cm$^2$,
in corresponding to the entropy distributions in Fig.1. For both
Models IV and V,  the $L_{\rm X}$-$T$ relations at 
$z=1$ are also shown (dashed lines). 
\label{fig3}}
\end{figure}

\clearpage 

\begin{figure}
\epsscale{0.5}
\plotone{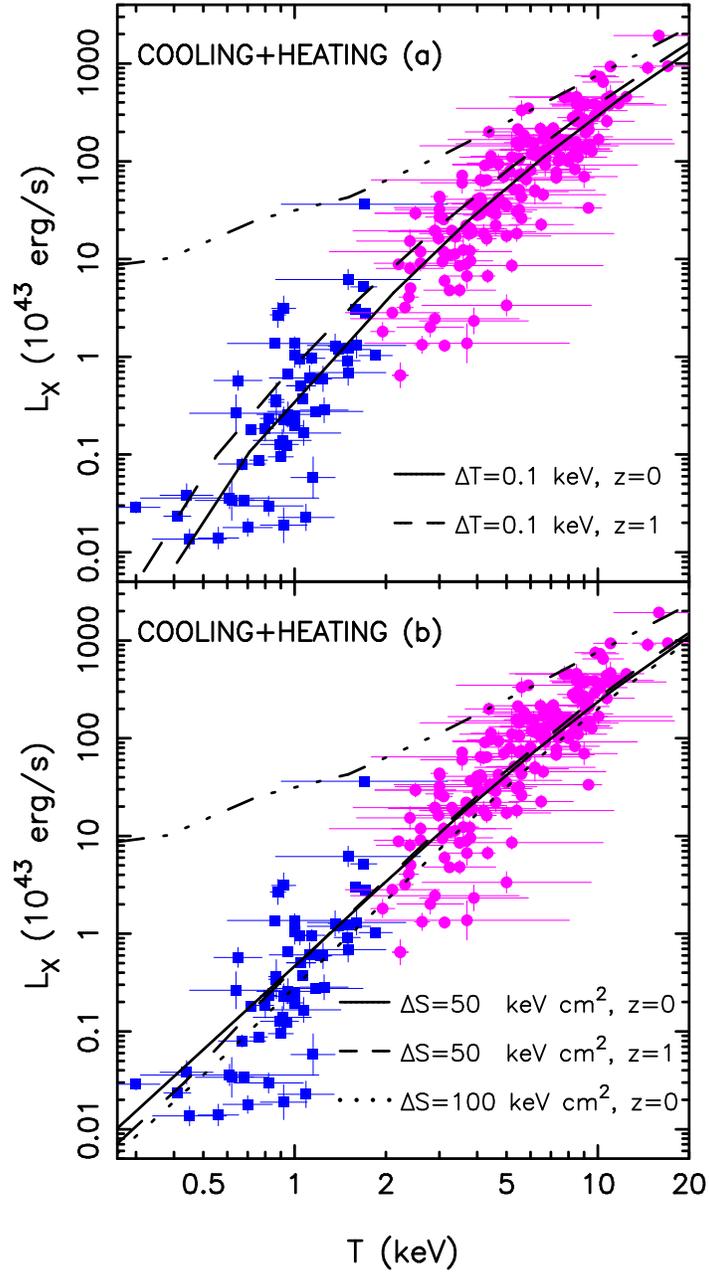}
\caption{The X-ray luminosity - temperature relations predicted by
cooling plus heating models VI (upper panel) and VII (lower panel). 
Dashed  lines are the  $L_{\rm X}$-$T$ relations at $z=1$.
\label{fig4}}
\end{figure}

\clearpage 

\begin{figure}
\epsscale{0.9}
\plotone{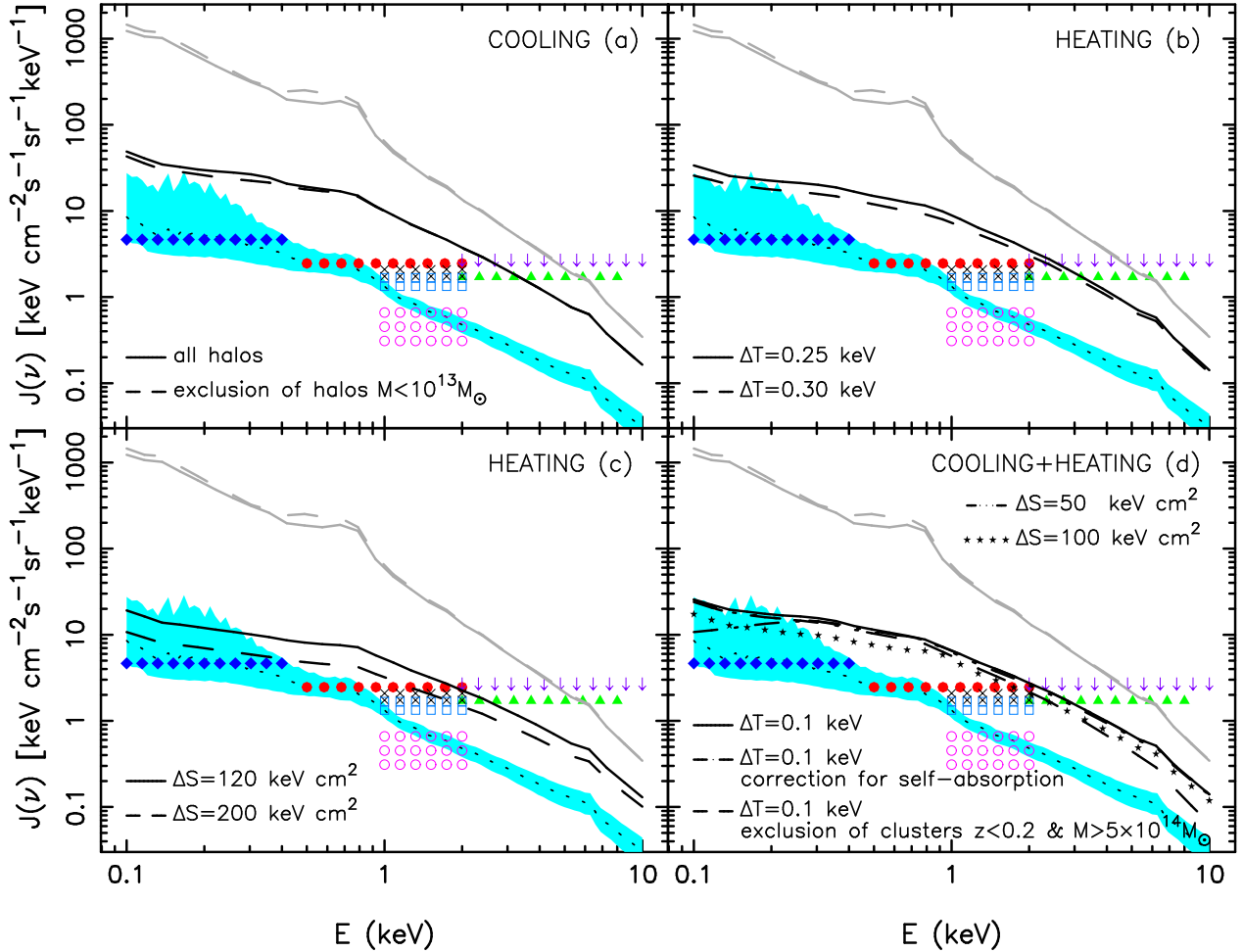}
\caption{Comparison of the predicted XRB with the observational 
upper limits on the contribution of diffuse IGM to XRB.   
Grey lines: the self-similar results with time-varying (solid) and constant 
(dashed) metallicity. Downarrows are the observational upper limits 
in the 2-10 keV band 
(Mushotzky et al. 2000; Giacconi et al. 2001; Tozzi et al. 2001). 
Filled triangles are the constraints obtained by Cowie et al. (2002)
in the 2-8 keV band.
Crosses, open squares and open circles 
in the 1-2 keV range correspond to three evaluations of 
the total XRB intensity by
Chen et al. (1997), Miyaji et al. (1998) and Gendreau (1997),
respectively, after the source contributions detected by
Hasinger et al. (1998), Mushotzky et al. (2000),  Giacconi et al. (2001)
\& Tozzi et al. (2001) 
are removed. Filled circles and diamonds are the upper limits derived by
Bryan \& Voit (2001) in the 0.5-2 and 0.1-0.4 keV bands, respectively. 
The total XRB estimated by summing up the contributions of groups
and clusters characterized by their X-ray luminosity functions and
the observationally determined $L_{\rm X}$-$T$ relations   
is shown by the shaded region (Wu \& Xue 2001).
\label{fig5}}
\end{figure}
\clearpage 

\begin{figure}
\epsscale{0.9}
\plotone{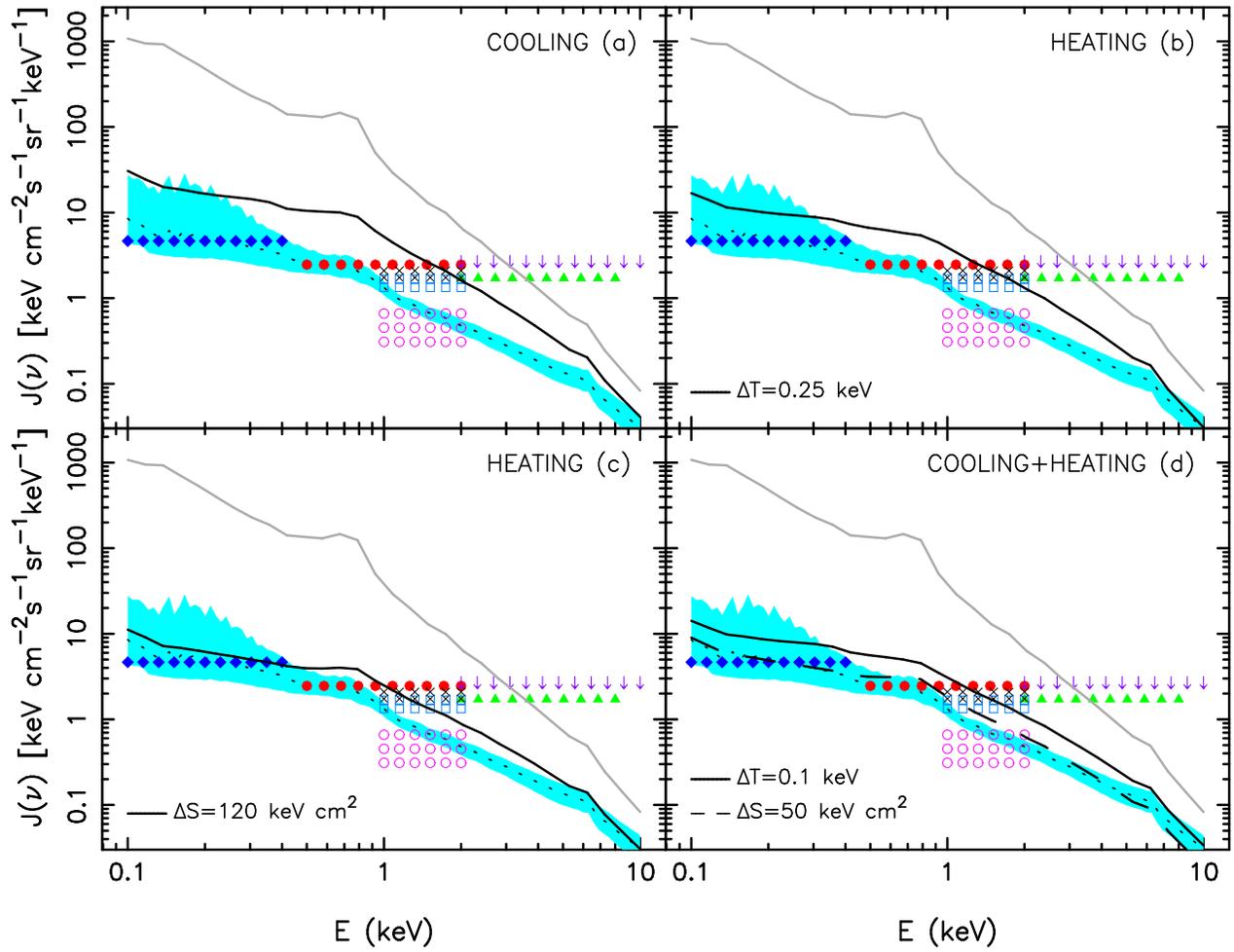}
\caption{The same as Fig.5 but for $\sigma_8=0.7$.
\label{fig6}}
\end{figure}

\end{document}